%% file: main.tex
\documentclass[
	aps, pra, twocolumn,superscriptaddress,
	floatfix, 
    nofootinbib,
	tightenlines,
	10pt
]{revtex4-1}
\usepackage[final]{graphicx}
\usepackage{times,bbm,amsmath,amssymb}
\usepackage{epsfig,color}
\usepackage{xcolor}
\usepackage{hyperref}
\hypersetup{
    colorlinks = true
}
\usepackage{cleveref}

\usepackage{float,siunitx}

\usepackage[caption = false]{subfig}

\usepackage[greek,english]{babel}
\usepackage{thumbpdf,enumerate}
\usepackage{booktabs}
\usepackage{sidecap}
\usepackage[scaled=.8]{couriers}    
\usepackage{pstricks}
\usepackage{multirow}
\usepackage{placeins}
 \usepackage{relsize}  
\usepackage{pst-grad,bm}
\usepackage{epigraph}
\usepackage{gensymb}
\usepackage{longtable}
\usepackage{booktabs}
\usepackage{gensymb}

\usepackage{soul}
\usepackage{ulem} 
\usepackage{acronym}

\usepackage{physics}

\usepackage{tikz}
\usepackage{subfiles}
\usepackage{braket}
\usepackage{acronym}
\usepackage{comment}

\usepackage[Symbol]{upgreek}
\usepackage[all=normal,floats=tight,mathspacing=tight,wordspacing=tight,paragraphs=normal,tracking=tight,charwidths=tight]{savetrees}
\everypar=\expandafter{\the\everypar\loosness=-1 }
\usepackage[style=american,autopunct=true]{csquotes} 
\usepackage{mathtools}

\begin{document}

\title{
{Reconfigurable continuously-coupled 3D photonic circuit for Boson Sampling experiments}}

\author{Francesco Hoch} 
\thanks{These two authors contributed equally}
\affiliation{Dipartimento di Fisica, Sapienza Universit\`{a} di Roma,
Piazzale Aldo Moro 5, I-00185 Roma, Italy}

\author{Simone Piacentini}
\thanks{These two authors contributed equally}
\affiliation{Dipartimento di Fisica, Politecnico di Milano, Piazza Leonardo da Vinci, 32, I-20133 Milano, Italy}
\affiliation{Istituto di Fotonica e Nanotecnologie, Consiglio Nazionale delle Ricerche (IFN-CNR), 
Piazza Leonardo da Vinci, 32, I-20133 Milano, Italy}

\author{Taira Giordani}
\affiliation{Dipartimento di Fisica, Sapienza Universit\`{a} di Roma,
Piazzale Aldo Moro 5, I-00185 Roma, Italy}

\author{Zhen-Nan Tian}
\affiliation{Istituto di Fotonica e Nanotecnologie, Consiglio Nazionale delle Ricerche (IFN-CNR), 
Piazza Leonardo da Vinci, 32, I-20133 Milano, Italy}

\author{Mariagrazia Iuliano}
\affiliation{Dipartimento di Fisica, Sapienza Universit\`{a} di Roma,
Piazzale Aldo Moro 5, I-00185 Roma, Italy}

\author{Chiara Esposito}
\affiliation{Dipartimento di Fisica, Sapienza Universit\`{a} di Roma,
Piazzale Aldo Moro 5, I-00185 Roma, Italy}

\author{Anita Camillini}
\affiliation{Dipartimento di Fisica, Sapienza Universit\`{a} di Roma,
Piazzale Aldo Moro 5, I-00185 Roma, Italy}

\author{Gonzalo Carvacho}
\affiliation{Dipartimento di Fisica, Sapienza Universit\`{a} di Roma,
Piazzale Aldo Moro 5, I-00185 Roma, Italy}

\author{Francesco Ceccarelli}
\affiliation{Istituto di Fotonica e Nanotecnologie, Consiglio Nazionale delle Ricerche (IFN-CNR), 
Piazza Leonardo da Vinci, 32, I-20133 Milano, Italy}

\author{Nicol\`o Spagnolo}
\affiliation{Dipartimento di Fisica, Sapienza Universit\`{a} di Roma,
Piazzale Aldo Moro 5, I-00185 Roma, Italy}

\author{Andrea Crespi}
\affiliation{Dipartimento di Fisica, Politecnico di Milano, Piazza Leonardo da Vinci, 32, I-20133 Milano, Italy}
\affiliation{Istituto di Fotonica e Nanotecnologie, Consiglio Nazionale delle Ricerche (IFN-CNR), 
Piazza Leonardo da Vinci, 32, I-20133 Milano, Italy}

\author{Fabio Sciarrino}
\email[Corresponding author: ]{fabio.sciarrino@uniroma1.it}
\affiliation{Dipartimento di Fisica, Sapienza Universit\`{a} di Roma,
Piazzale Aldo Moro 5, I-00185 Roma, Italy}

\author{Roberto Osellame}
\email[Corresponding author: ]{roberto.osellame@cnr.it}
\affiliation{Istituto di Fotonica e Nanotecnologie, Consiglio Nazionale delle Ricerche (IFN-CNR), 
Piazza Leonardo da Vinci, 32, I-20133 Milano, Italy}

\begin{abstract}
Boson Sampling is a computational paradigm representing one of the most viable and pursued approaches to demonstrate the regime of quantum advantage. Recent results have demonstrated significant technological leaps in single-photon generation and detection, leading to progressively larger experimental instances of Boson Sampling experiments in different photonic systems. However, a crucial requirement for a fully-fledged platform solving this problem is the capability of implementing large-scale interferometers, that must simultaneously exhibit low losses, high degree of reconfigurability and the realization of arbitrary transformations. In this work, we move a step forward in this direction by demonstrating the adoption of a compact and reconfigurable 3D-integrated platform for photonic Boson Sampling. We perform 3- and 4-photon experiments by using such platform, showing the possibility of programming the circuit to implement a large number of unitary transformations. These results show that such compact and highly-reconfigurable layout can be scaled up to experiments with larger number of photons and modes, and can provide a viable direction for hybrid computing with photonic processors.
\end{abstract}

\maketitle

Since the original proposal of a computational paradigm based on the rules of quantum mechanics \cite{Feynman1982}, large research efforts have been devoted to identifying the optimal approach to implement a universal quantum computer \cite{DiVincenzo_criteria,chuang2010}. Besides the great promises provided by such computational paradigm, the implementation of a large-scale universal quantum device capable 
to outperform or, at least, to be comparable with a classical computer is still a challenging task. In view of the very recent advances in quantum technologies, approaching the noisy intermediate-scale quantum (NISQ) era, the current 
target is to reach 
a fundamental milestone named quantum computational advantage. The goal is to achieve, unambiguously and possibly with different platforms and approaches, the scenario where a quantum device is capable of solving a specific task faster than any classical counterpart. 

\begin{figure*}[ht!]
\centering
\includegraphics[width=0.99\textwidth]{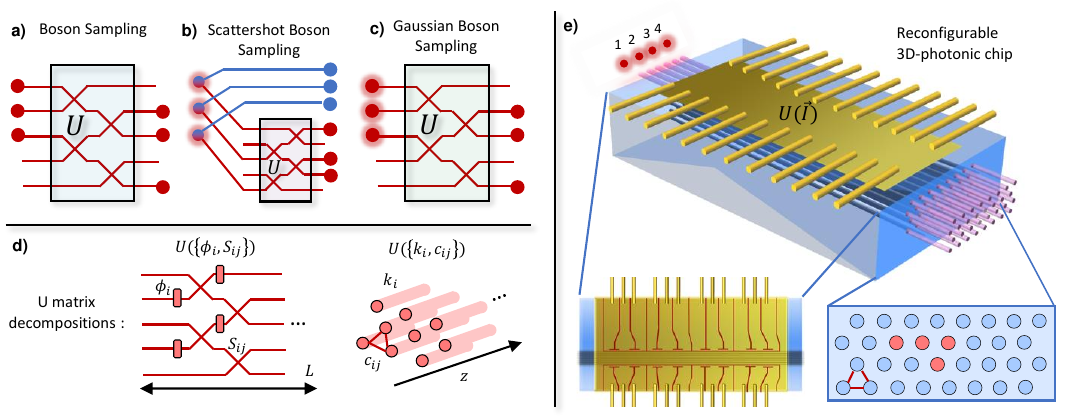}
\caption{\textbf{Boson Sampling in a 3D continuous-coupling integrated device. a)} Boson Sampling (BS) and the most recent variants, \textbf{b)} Scattershot Boson Sampling (SBS) and \textbf{c)} Gaussian Boson Sampling GBS). The corresponding computational problems require sampling from the output distribution using different input quantum states of light, such as Fock states in BS, two-mode squeezed vacuum states in SBS and single-mode squeezed vacuum states in GBS. The common element among the schemes is the optical random circuit, described by the unitary evolution $U$. \textbf{d)} The most widely adopted decomposition of the operator $U$ is via a network of beamsplitters, with splitting ratios $S_{ij}$, and phase-shifts $\phi_{i}$ (left); an alternative implementation exploits continuous-coupling by evanescent waves among waveguides (right) depending on the coupling coefficients $c_{ij}$ and the propagation constants $k_i$, where both may vary along the direction $z$. \textbf{e)} Overview of the reconfigurable 3D integrated photonic chip, realized through the femtosecond laser writing technique. The device is composed by 32 optical-modes arranged in a triangular lattice, as showed in the inset reporting the transverse section of the sample. In red we have highlighted the input modes employed in the 3- and 4-photon experiment. The transformation $U$ is controlled by the 16 resistors fabricated on top of the glass sample. The second inset shows the top view of the electrical circuits that controls the currents  $I_i$ applied to the resistors. 
}
\label{fig:chip}
\end{figure*}

Among the several quantum algorithms that provide a computational speed-up, two strategies emerge as the most suitable for their experimental realizations \cite{Harrow2017_supremacy}. The first one requires to sample the output states produced by a random quantum circuit. This paradigm has been recently implemented in quantum processor{s} {based on} superconducting qubits \cite{Arute2019, Wu_qubits2021}. The second approach is based on a different albeit related task, named Boson Sampling (BS) \cite{AA,Brod19review}, which requires sampling from the distribution of non-interacting bosons scattered by a random unitary transformation. Here, the hardness of simulating such a bosonic system is strictly connected to quantum interference effects due only to particle indistinguishability \cite{HOM}. Classical simulation of Boson Sampling, even approximately, is computationally-hard since it requires calculation of permanents of complex-entried matrices, a \#P-hard problem. A natural way to reproduce this dynamics and sampling effectively from the output distribution of such process can be obtained via a photonic quantum processor. Reaching the quantum advantage regime with this approach requires the generation of a set of $n$ highly-indistinguishable single photons, which evolve via quantum interference in a low-loss linear optical network with $m \sim n^2$ modes capable of implementing a random transformation according to the Haar measure (see Fig.~\ref{fig:chip}a). Samples are then collected by direct measurement of the output modes. In this direction, several proof-of-principle implementations have been reported exploiting photonic platforms {\cite{Broome13boson,Spring13boson,Crespi13boson,Tillmann13boson,Loredo17,He_Pan17,Wang_Pan17,Wang_2018,zhong2018scattershot, gao2019experimental}}, including one of the latest experiments with the detection of 14 photons after propagation in an interferometer with 60 ports \cite{Pan_20photons}.

Starting from these results, research efforts have been dedicated to technological advances enabling to enlarge the dimensionality of photonic processors, and to theoretical investigations \cite{Neville2017, CC} aimed at a precise definition of the limits of a classical simulation of Boson Sampling with experimental imperfections. In parallel, the Boson Sampling paradigm has triggered the definition of 
a set of variants of the original formulation with the aim of improving the efficiency of the corresponding photonic platform while preserving the classical computational complexity of the task. Examples include (but are not limited to) Scattershot Boson Sampling (SBS) \cite{Lund_SBS} and Gaussian Boson Sampling (GBS) \cite{HamiltonGBS}
. SBS exploits probabilistic parametric down-conversion sources placed at each input mode of the interferometer in a heralded configuration (see Fig.~\ref{fig:chip}b), leading to an exponential growth in the samples acquisition \cite{Bentivegna15, zhong2018scattershot,Paesani2019}. In the GBS paradigm  \cite{Paesani2019,Zhong19,Arrazola2021}, Fock states are replaced by single-mode squeezed vacuum states (see Fig.~\ref{fig:chip}c). By using the GBS approach, recent experiments have reported the achievement of quantum advantage with a photonic platform \cite{Zhong_GBS_supremacy, Zhong_phase_2021}. Furthermore, GBS has been recently pointed out to have potential application for hybrid classical-quantum computing, due to the connection with other problems including graph theory \cite{Arrazzola_densesubgraph, Shuld_GBS_graphsimilarity} or simulation of molecular vibronic spectra \cite{Huh2015_vibronic, Banchi_vibronic}. In all previous experiments reporting Boson Sampling instances, including its variants, different platforms have been employed to realize linear optical transformations. However, all the requirements for a fully-developed processor, namely low-loss, reconfigurability, and the possibility to implement random transformations, are currently not fulfilled simultaneously in a single system. {Indeed, low-loss systems are necessary to avoid spoiling the complexity of the process \cite{Qi_NoisyGBS,GarciaPatron2019simulatingboson}, and full-reconfigurability is a crucial requirement on two main aspects. On one side, such a feature is essential to benchmark the effective Haar-randomness of the platform, which is at the basis of complexity conjectures \cite{AA, DetailedstudyGBS}. On the other side, reconfigurability is needed to achieve programmable processors for applications beyond the quantum advantage demonstration \cite{Arrazola2021}.} {With current-up-to-date experiments,} low-loss platforms lack an active reconfiguration capability to change the transformation implemented by the optical circuit {\cite{Pan_20photons,Zhong_GBS_supremacy, Zhong_phase_2021}} {and their capability of Haar-random operations has not been demonstrated yet}. Conversely,  integrated photonic architectures based on universal decompositions \cite{Reck_1994,Clements:16} permit full reconfiguration capabilities, but still require further technological improvements also in terms of loss-reduction to scale up the number of modes.

Here, we perform a step forward towards developing a photonic platform encompassing all the aforementioned characteristics. We report the realization of a photonic reconfigurable integrated device with a compact structure. Such device provides significant advantages in terms of losses and number of modes{. In addition, it} possesses a high degree of reconfigurability enabling the implementation of a large number of transformations{, in contrast to previous works that implement static integrated circuits via the same technology \cite{gao2019experimental, jiao2020twodimensional}}. This architecture takes advantage from the 3D-capability of the femtosecond laser-writing technique of waveguides in glass \cite{Gattass2008, Wang2020_review}. The optical circuit is realized via continuous-coupling of 32 waveguides arranged in a triangular lattice. The device is then controlled by changing the currents applied to the 16 heaters fabricated on top of the interferometer. We show the reconfiguration capabilities of the device by providing a thorough analysis on the set of unitary transformations which can be reached by the system. Then, we benchmark our platform in the Boson Sampling scenario by performing and validating 3-photon and 4-photon experiments with several different configurations of the optical transformation.
\section*{Results}
\subsection*{Reconfigurable integrated 3D photonic chip}

\begin{figure*}[ht!]
\centering
\includegraphics[width=\textwidth]{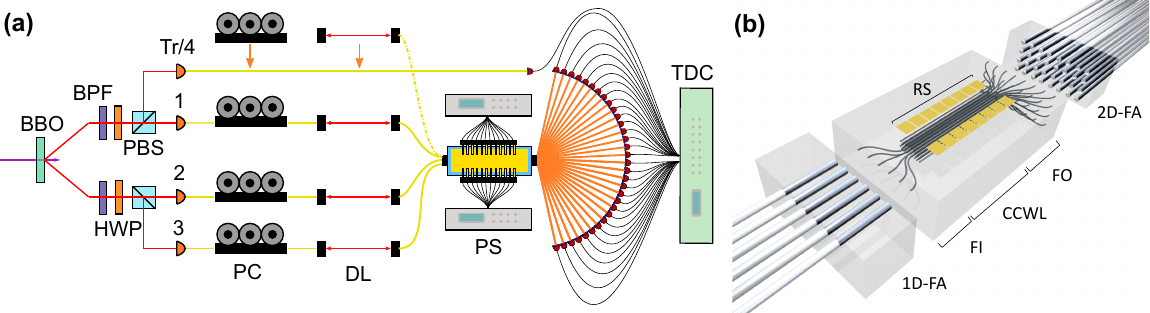}
\caption{\textbf{Scheme of the experimental apparatus. a)} A parametric down-conversion 
process in a Beta-Barium Borate crystal 
generates one- and two-pair photon states. Generation in the single-pair regime is employed for the unitary reconstruction procedures, while 3- and 4-photon states are employed for experiments in the Boson Sampling framework. Photons are prepared in their polarization and temporal degrees of freedom before coupling in the input single-mode fiber array. After evolution, photons are finally detected via a set of $32$ single-photon avalanche photodiodes connected to a $32$-channel time-to-digital converter for the reconstruction of the coincidence pattern. \textbf{b)} Schematic of the in- and out-coupling of the single photons with the 3D photonic circuit: one- and two-dimensional fiber arrays connect to the fan-in and fan-out sections of the circuit.
\textbf{Legend:} BBO - beta-barium borate crystal, BPF - band-pass filter, HWP - half-wave plate,  PBS - polarizing beamsplitter, PC - polarization controller, DL - delay line, PS - power supply, TDC - time-to-digital converter, 1D-/2D-FA - one-/two-dimensional fiber array, FI - fan-in, FO - fan-out, CCWL - continuously-coupled waveguide lattice, RS - resistors.
}
\label{fig:apparato}
\end{figure*}

Integrated photonics has demonstrated significant advances in realizing complex optical circuits for quantum information processing \cite{Wang2020_review}. Nowadays integrated optical circuits are the most promising photonic platforms to implement large-scale interferometer with high-level of reconfigurability. The unitary operations that describe a given optical circuit are often decomposed in elementary optical units. The scheme by Reck et al. \cite{Reck_1994} decomposes an optical circuit with $m$ optical modes in the product of $m(m-1)/2$ two-mode beamsplitters and single-mode phase-shifters. A more recent algorithm by Clements et al. \cite{Clements:16} optimizes the arrangement of the optical elements to minimize the sensitivity to fabrication imperfections and to photon losses. In this framework, any unitary transformation is determined by the set of phases $\phi_{i}$ and splitting ratios $S_{ij}$ in the circuit (see Fig.~\ref{fig:chip}d). In reconfigurable integrated circuit, heaters placed in correspondence to the waveguide enable to change locally the refractive index of the material thus inducing a change in the relative phases among the optical modes. Despite the aforementioned schemes have been adopted in several experiments \cite{Crespi13boson, Carolan15, harris2015bosonic, wangpaesani, Taballione2019, Arrazola2021}, scaling the circuit to large number of modes is a challenging task in terms of size of the device, related to the amount of losses and to the number of required optical elements.  In Fig.~\ref{fig:chip}d we depict 
the approach that we employ in this paper, which is different with respect to the traditional decomposition of unitaries in beamsplitters and phase-shifters.
Such a scheme exploits the continuous coupling of the radiation in waveguides arrays. In this framework it is possible to associate a Hamiltonian $\mathcal{H}(z)$ to the system given the coupling coefficients  $c_{ij}(z)$ between neighbouring modes and the propagation constants of the modes $k_i(z)$. Coupling coefficients and propagation constants may change along the propagation coordinate $z$. 
The unitary transformation of the circuit can be obtained by integrating  $\mathcal{H}(z)$ for the whole length of the interaction region, which maps the time evolution of the system. In particular, in our case we employ a three-dimensional array, in which waveguides are arranged according to a triangular-lattice cross-section. Such architecture offers practical advantages in terms of compactness, losses and circuit length in comparison to other approaches. 
Exemplarily, it is striking to note that in this work we have achieved 
reconfigurable 32-modes random unitaries in a 7.5-cm long optical chip, while a reconfigurable discretely-coupled interferometer with the same number of modes and 
produced with the same technology would have required a device length of about 30 cm.
Such aspects are widely discussed in Supplementary Note 1. 

In detail, our integrated device, depicted in Fig.~\ref{fig:chip}e, comprises 32 continuously-coupled single-mode waveguides in a $8\times 4$ arrangement, fabricated by femtosecond laser direct writing in a borosilicate glass substrate \cite{Gattass2008,arriola2013}. 
The waveguide positions are randomly modulated along the propagation coordinate $z$ with respect to the positions of a regular triangular lattice. These modulations introduce randomness in the $c_{ij}$ coefficients and thus in the circuit transformation, which otherwise would be highly symmetric and not suitable for a Boson Sampling experiment. 
To add circuit reconfigurability, 16 resistive heaters have been patterned, by femtosecond laser ablation, on a gold film deposited on the substrate surface. The resistors are equally distributed on the two sides of the interaction region. An external power supply controls the currents applied to the resistors. Dissipated power in this process induces thermal gradients in the substrate and thus locally changes  the refractive index of the waveguides \cite{Flamini2015,Pentangelo2021}. This modulates the propagation constants $k_i$ of the waveguides by the thermo-optic effect, thus allowing to change dynamically the transformation $U$ implemented in the integrated device. 
Hence, we demonstrate that the thermo-optic phase-shifting technology can also be applied to three-dimensional, continuously-coupled waveguide devices. Additional details on the circuit geometry and on the fabrication process are provided in the Methods section.

\subsection*{Experimental platform}
\label{sec:exp}

\begin{figure*}[ht]
\centering
\includegraphics[width=0.99\textwidth]{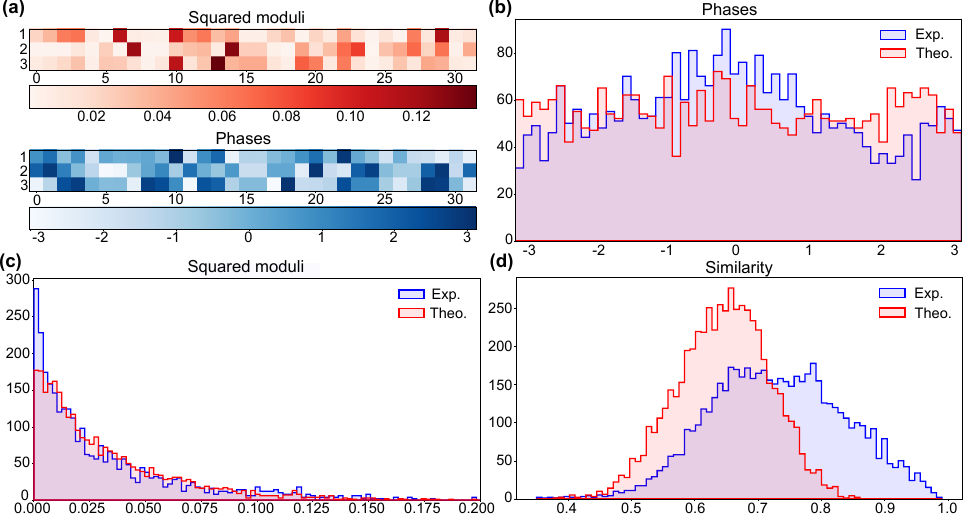}
\caption{
\textbf{Experimental reconstruction of the $[3\times 32]$ sub-matrix and comparison with the Haar-random matrices.} \textbf{(a)} Experimental reconstruction of the squared moduli $\rho_{ij}^2$ (red) and phases $\theta_{ij}$ (blue) for three input ports of the reconfigurable photonic chip, highlighted by the labels in the figure. The chip is set on a random configuration of currents. Each input port represents a row of the unitary transformation applied to the input state. \textbf{(b)-(c)} Comparison of the phases and the squared moduli frequency distribution respectively, between 15 experimentally reconstructed sub-matrix (blue) and 15 $[3\times 32]$ sub-matrix sampled from the Haar-random unitaries (red).
\textbf{(d)} Distribution of the similarity between the squared moduli of two columns of different sub-matrix. We repeated the measurement for $\sim$ 200 different configurations of the currents in the chip. In red it is shown the theoretical similarity distribution obtained by sampling columns distributed according to the Haar measure. The overlap between the two histograms is the {62.4\%} of the total area. 
}
\label{fig:frequency}
\end{figure*}

Let us now illustrate the components of the experimental platform employed to benchmark the photonic device and to collect the 3- and 4-photon samples (see Fig.~\ref{fig:apparato}a).
We exploit one- and two-pair emission in a type-II spontaneous parametric down-conversion source composed of a beta-barium borate crystal operating at \SI{785}{\nano \metre}. The first stage of the apparatus includes all the optical components to generate the state resource to perform sampling with either indistinguishable and distinguishable photons. Photons spectra are filtered through a $\SI{3}{\nano \metre}$ band-pass filter. Then, photons are split in four different spatial modes according to the polarization via half-wave plates and polarizing beamsplitters, and coupled into single-mode fibers. Photons are controlled in polarization and in time-of-arrival by polarization controllers and delay lines respectively, in order to tune their degree of indistinguishability. Then, they are injected into the reconfigurable integrated chip via an array of 6 single-mode fibers that has been aligned and glued to the device. A fan-in waveguide section leads the photons to selected inputs of the waveguide array (see Fig.~\ref{fig:apparato}b). After the evolution in the integrated device, a fan-out waveguide section leads the photons to a $8\times4$ rectangular multimode fiber array that matches the fan-out geometry. It is worth noting that the 2D fiber array further helps the compactness of the device by greatly reducing the length of the fan-out section (see Supplementary Note 1 and Supplementary Figures 5-6).
The detection stage includes 32 single-photon avalanche photo-diodes. We have developed a custom software that simultaneously controls the delay lines, the power supply and the 32-channel time-to-digital converter module to record two- and four-fold coincidences. This implies a full control over 
the unitary transformation implemented in the circuit, the switching 
between indistinguishable and distinguishable photons, and the recording and processing of the 
data samples.

\subsection*{Unitary matrix sampling and reconstruction}

Control over the unitary transformations implemented in the chip is performed by tuning the currents in the resistors. To verify the classes of matrices $U$ that can be implemented by the device due to its reconfigurability, we have reconstructed a large noumber of different evolutions, each corresponding to a different setting for the currents in the resistors. Thus, a crucial ingredient was the adoption of a fast and efficient reconstruction algorithm. Additionally, an efficient and fast reconstruction of the unitary transformations is a fundamental step also for benchmarking and validating the 3- and 4-photon experiments described in the next section.  We made use of an adapted version of the method reported in \cite{laing2012superstable} (see Supplementary Note 2 and Supplementary Figure 7). The latter envisages the measurements of two-photon Hong-Ou-Mandel (HOM) dips resulting from pairs of photons injected in different combinations of input ports. In our case we restricted the measurements to two of the possible input pair combinations for the reconstruction of $3\times32$ sub-matrix and to only three pairs in the case of $4\times32$ sub-matrix.  From each input pair we analysed $496$ HOM dips, namely all the possible non-redundant pairs obtained by the combination of the 32 output ports. From these measurements we have extracted the information about the moduli $\rho_{ij}$ and the phases $\theta_{ij}$ of the sub-matrix elements expressed as $U_{ij}=\rho_{ij} e^{i \theta_{ij}}$. In Fig.~\ref{fig:frequency}a we have reported an example regarding the 96 squared moduli and phases of one of the 15 different $3 \times 32$ sub-matrices reconstructed in this work. Our next step was to prove that the random unitaries, sampled by changing the currents configuration in the circuit, were drawn from a distribution as close as possible to the Haar measure. This requirement is fundamental to ensure the hardness of BS. To this aim we compared the distributions of phases and moduli of the 15 sub-matrix measured in the experiment with the one retrieved from likewise Haar-random extracted unitary matrices (Fig.~\ref{fig:frequency}b-c). In both cases we have obtained a good agreement between the two distributions. The slight deviations from the theoretical histograms can be explained by experimental artifacts related to losses, imperfections in the apparatus and to the algorithm for the reconstruction of the matrix. We discuss these effects in Supplementary Note 3 and Supplementary Figures 8-10. As a final benchmark of the device, we measured 200 columns of different unitaries and calculated the similarities between the distributions given by the squared moduli of each column. The measurements were performed by sending one photon in the device and measuring it at the output in coincidence with his correlated one in a two-photon experiment. The unitaries have been generated by a uniform sampling of the electrical power dissipated in the resistors. Also in this analysis we find a  good agreement with expectations, signified by the overlap with the histogram of the similarities calculated from the columns of Haar-random matrices shown in Fig.~\ref{fig:frequency}d.
The latter result represents one of the first investigations on the level of randomness that can be reached in this continuously-coupled waveguide architecture by changing only the propagation constants via the thermo-optic effect. The {similarity to the Haar-random distribution} could be improved by engineering the sampling strategy of the dissipated powers, which here have been extracted from a uniform distribution. Note that{, in the discrete-decomposition schemes (Reck, Clements)} there exist algorithms to set the optical circuit to sample from the Haar distribution. However, the phase shift values and beamsplitter reflectivitie do not display trivial distributions \cite{Russel_2017_Haar_Reck, burgwal2017_Haar_Clements} which in turn require {complex} settings of the external control circuit. More precisely, by increasing the dimension of the matrix, the parameter distributions tend to be more and more peaked. For example, the uniform sampling of the dissipated electrical powers employed in this work is far from the correct sampling to generate random Haar matrices in discrete optical circuits. An exhaustive and conclusive answer regarding the possibility to extract matrices from a distribution closer to the Haar measure with {a} continuously-coupled {waveguide} architecture needs further studies both from a theoretical and experimental point of view.

\subsection*{Experimental Boson Sampling in a 3D reconfigurable circuit}

\begin{figure*}[t]
\centering
\includegraphics[width = 0.99\textwidth]{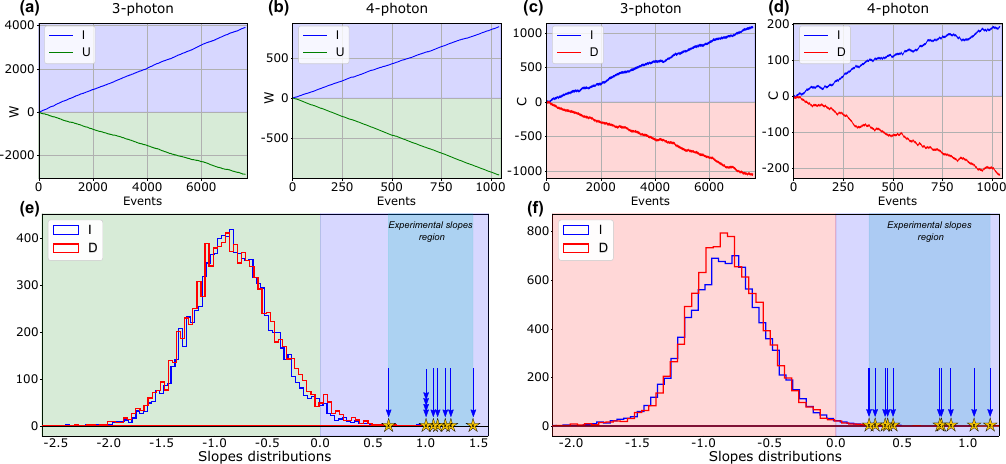}
\caption{\textbf{Boson Sampling data.} Validation of the 3-photon (blue points in \textbf{(a)}) and 4-photon (blue points in \textbf{(b)}) Boson Sampling experiments against the uniform sampler hypothesis (green point in \textbf{a-b}). 
\textbf{(c-d)} Validation against the distinguishable photons sampler of 3- and 4-photon events. The test is applied using the same reconstructed sub-matrix in \textbf{(a)} and \textbf{(b)}. The distinguishable samples (red points) were collected by adjusting the relative time delay between the input photons.
\textbf{(e)} Histograms of simulated 3-photon slopes for counter $W$ (normalized to the slopes of distinguishable particles data) in the case of validation against the uniform sampler using a random extracted $U$ transformation, i.e. $U$ matrices that do not match the actual operation implemented in the circuit.
The obtained slopes of the 10 different 3-photon BS experiments are highlighted by the arrows (and by the corresponding stars). The experimental points were validated with the reconstructed sub-matrix retrieved from the two-photon reconstruction. All the experimental slopes values are in the positive range and far from the histograms average, thus showing the correct validation of the performed experiment. In \textbf{(f)}, we performed the same analysis for the validation against the distinguishable photons hypothesis. 
}
\label{fig:validations}
\end{figure*}

After characterization of the device, we have then performed 3- and 4-photon experiments in the Boson Sampling framework with our integrated system. The 4-photon state from double-pair emission generated by the source can be written as $\rho^{\mathrm{in}} \sim \alpha \ket{1111}\bra{1111} + \beta \ket{2002}\bra{2002} + \gamma \ket{0220}\bra{0220}$ (see Supplementary Note 4 and Supplementary Figure 11), by expressing the density matrix in the occupation number of the four modes and neglecting the higher order of multi-pair emission. To inject a 3-photon $\vert 111 \rangle$ input state, one of the four output modes of the source is directly measured and thus acts as a trigger. To this end, we have discarded one output mode of the chip, due to the requirement of using one detector for the trigger photon. Four-fold coincidences between the trigger photon and three output modes are then recorded, providing the output samples.  In the 4-photon experiment, we have sampled from the entire $\rho^{\mathrm{in}}$ by directly connecting the four output modes of the source to the integrated device (see Fig.~\ref{fig:apparato}). In this case, we have sampled from all the output ports of the device. For all reported experiments, our measurements are restricted to the collision-free events. Such choice does not affect significantly the outcomes of the experiment since the configurations with more than one photon per mode display very low probabilities in the regime where the number of photons is much smaller compared to the number of the optical modes \cite{Spagnolo13birthday}. 

Let us now illustrate the analysis of the experimental samples collected in a 3- and 4-photon BS routines. In this context, the problem of data validation is pivotal to assess the correctness of the sampling process, especially in the regime in which it is not possible to reproduce the output of the experiment with classical resources. In the past years several tests have been developed to rule out classical models, such as the uniform and distinguishable particle samplers, that could reproduce some features of the BS output distribution \cite{Aaronson14, Spagnolo14,Carolan14, Crespi16, Walschaers16,Viggianiello18, agresti2019pattern, Giordani18, FlaminiTSNE, Giordani_2020}. In Fig.~\ref{fig:validations} a-b we report one instance of the validation of a 3- and 4-photon BS against the uniform distribution test \cite{Aaronson14}, assuming the input states described above for the two scenarios. We also employ the same data to validate the experiment against the distinguishable particles hypothesis \cite{Spagnolo14} (see Fig.~\ref{fig:validations} c-d, Supplementary Note 5 and Supplementary Figures 12-15). In both tests each event collected in the experiment increases or decreases a counter, $W$ for the case of uniform sampler and $C$ for the distinguishable particle sampler, according to a likelihood ratio test. Positive slopes are the signatures of the successful validation of the data. These kinds of hypothesis tests require a good modelling of the system, including the knowledge of the input state and of the unitary transformation applied to the state. The latter has been reconstructed through two-photon measurements and exploiting the reconstruction algorithm discussed above (see also Supplementary Note 2). We have performed 10 different 3-photon BS experiments, and 3 different 4-photon with likewise configurations of the optical circuit. All the data were successfully validated against the two hypotheses. In Fig.~\ref{fig:validations} e-f we underline the sensitivity of these validation tests to the reconstruction of the matrix representing the optical circuit. The histograms report the distribution of the slopes normalized to the one of distinguishable particles when the unitary is chosen randomly and does not coincide with the actual transformation performed in the circuit. We note that, in absence of correspondence between the unitary transformation related to the data and the one related to the likelihood ratio computations, these tests assign the data to the negative hypothesis independently from the particle statistics. In the same figure we report the experimental slopes describing the set of 10 different 3-photon BS validated in this work using the unitary matrices reconstructed via our algorithm. The experimental points (stars in the Fig.~\ref{fig:validations}e-f) are distant more than 3 standard deviations with respect to the average of the histograms, representing validations with random unitaries that do not correspond to the circuit from which the samples were generated as described above. Such additional result reinforce the successful validation of the performed experiments, and benchmark the reconstruction accuracy of the optical circuit showing a high degree of control of the platform. 

\section*{Discussion}

In this article we have reported on the implementation and benchmarking of an integrated platform, with a compact 3D layout based on continuous waveguide coupling which includes a set of heaters to enable a high degree of reconfigurability. The employed architecture, achievable by exploiting the unique capabilities of the femtosecond laser {micromachining} technique, can be scaled up to larger number of modes. Indeed, we have successfully shown that our 32-mode device can implement a relevant portion of transformations according to the Haar measure, a fundamental requirement to fulfil the randomness hypothesis at the basis of the complexity of Boson Sampling. We have then implemented and validated 3- and 4-photon experiments, showing the viability of using such platform for future large-scale Boson Sampling instances. {In fact, the devices fabricated with femtosecond laser technique can  be  interfaced  with  other  types  of  single-photon  sources, such as deterministic quantum dot emitters \cite{Anton:19}. Deterministic sources are the most promising for scaling up the number of indistinguishable photons in a genuine Fock state \cite{Somaschi2016, Wang2019}. The same waveguide{} fabrication technology is able to realize integrated parametric sources \cite{Atzeni:18} that could be included in our device to mitigate the current losses in {coupling photons to single-mode fibers and at fiber-waveguide interfaces}.  Furthermore, the scheme with arrays of integrated sources could be feasible for realizing Scattershot and Gaussian Boson Sampling variants in a fully integrated platform \cite{Paesani2019, Arrazola2021}.} The capability of devising a fully-reprogrammable, large scale and compact photonic processor is fundamental also to envisage future applications beyond the original scope of the Boson Sampling. {To this aim, {future studies shall} focus on methods that {allow one} to control and program the transformations that the device can implement. This means {finding} a model that links the parameters of the unitary transformations {to} the dissipated electrical powers in the heaters, which {} is a problem {still not addressed in the} literature {for reconfigurable continuously-coupled waveguide circuits}. We {believe that} the use of a black-box approach, via neural networks and optimization algorithms  {\cite{Youssry2020,Skryabin2021}, is possible and promising}. Further improvements in the reconfigurability could be brought from modifications of the present device architecture, {e.g. by increasing} the number of heaters and {by changing} their arrangement with respect to the waveguide positions}.

{Given the aforementioned potentiality and advantages reported in this work} such a platform is expected to be at the basis of recent proposals of hybrid computing architectures  \cite{Pointpro, 
ArrazolaQOpt, Arrazzola_densesubgraph, Shuld_GBS_graphsimilarity,Huh2015_vibronic, Banchi_vibronic}.

\section*{Methods}

\textbf{3D photonic circuit.} The circuit consists of a triangular lattice with 32 continuously-coupled waveguides characterized by an average pitch of $\SI{11}{\micro\metre}$ (see Fig.~\ref{fig:chip}e). Indeed, each waveguide is shifted from its standard position of a quantity randomly chosen between $0$ and $\SI{2}{\micro\metre}$ along a random direction. Such shifts are varied with continuity along the whole coupling region, which is overall 36~mm long. 
In order to allow the coupling of single photons into the circuit, 6 central waveguides of the array extend through a fan-in region, rearranged in a single row with spacing of $\SI{127}{\micro\metre}$. At the output side, a fan-out expands all the waveguides in a $8 \times 4$ rectangular lattice with a pitch of $\SI{250}{\micro\metre}$. Photons are coupled to the input of the circuit with a linear single-mode fiber array, while are collected at the output by a rectangular multimode fiber array (see Fig.~\ref{fig:apparato}b). Input and output fiber arrays match the geometry of the fan-in and fan-out regions, respectively. Total insertion losses are $3.5 \pm \SI{0.1}{\deci \bel}$, depending on the input waveguide considered.

\textbf{Fabrication process.} The 3D photonic circuit was fabricated by femtosecond laser writing in a boro-aluminosilicate glass substrate (EAGLE XG, Corning) extending on an area of $75 \times \SI{12}{\milli\metre\squared} $. The laser source (PHAROS, Light Conversion), operating at a wavelength of $\SI{1030}{nm}$, was configured to produce pulses with duration of $\SI{170}{\femto \second}$ and energy equal to $\SI{290}{\nano \joule}$ at a repetition rate of $\SI{1}{\mega \hertz}$. The laser was focused with a 20$\times$ ($NA=0.5$) water-immersion objective, while the substrate was translated at $\SI{20}{\milli \metre \per \second}$ for six consecutive scans. To ensure efficient reconfigurability, the circuit was inscribed at $\SI{30}{\micro \metre}$ from the surface. After a thermal annealing step \cite{Corrielli2018_polins}, single-mode waveguides operating at $\SI{785}{\nano \metre}$ with a $1/e^2$ mode diameter of $\SI{4.5}{\micro \metre}$ were obtained. Heaters were fabricated by depositing gold on the chip surface and patterning the electrical circuit with the process reported in \cite{Ceccarelli2019}. A total number of 16 resistors (length $\SI{3}{mm}$, resistance $70 \pm \SI{13}{\ohm}$) were arranged in two parallel rows at the sides of the coupling region. To guarantee proper heat dissipation, the device was mounted on an aluminium heat sink. \\

\section*{Data availability}
The data that support the findings of this study are available from the corresponding authors upon reasonable request.

\section*{Acknowledgments} 
This work is supported by the European Union’s Horizon 2020 research and innovation program through the FET project PHOQUSING (“PHOtonic Quantum SamplING machine” - Grant Agreement No. 899544) and under the ERC project CAPABLE ("Composite integrAted Photonic plAtform By ultrafast LasEr micromachining" - Grant Agreement No. 742745). The authors wish to acknowledge financial support also by MIUR (Ministero dell’Istruzione, dell’Università e della Ricerca) via project PRIN 2017 “Taming complexity via QUantum Strategies: a Hybrid Integrated Photonic approach” (QUSHIP - Id. 2017SRNBRK). Z.-N. T. acknowlegdes funding by Quantera programme (project HiPhoP - High-dimensional quantum Photonic Platform; grant agreement no. 731473). Fabrication of the device was partially performed at PoliFAB, the micro- and nanofabrication facility of Politecnico di Milano (www.polifab.polimi.it). The authors would like to thank the PoliFAB staff for valuable technical support.

\section*{Competing Interests}
The authors declare they have no competing interests.

\section*{Author contributions} 
F.H. and S.P. contributed equally to this work. F.H., T.G., G.C., N.S., F.S., S.P., A.Cr., and R.O. conceived the experiment. S.P., Z.-N.T., F.C., A.Cr., and R.O. fabricated and characterized the integrated device using classical optics. F.H., T.G., M.I., C.E., A.Ca., G.C., N.S., and F.S. carried out the quantum experiments and performed the data analysis. All the authors discussed the results and contributed to the writing of the paper.

\newpage

\subfile{main_sm}

\newpage

\bibliography{bibliography}

\end{document}

%% file: main_sm.tex
\onecolumngrid
\begin{widetext}
\section*{Supplementary Information: {Reconfigurable continuously-coupled 3D photonic circuit for Boson Sampling experiments}}
\twocolumngrid
\end{widetext}



\section{Footprint of integrated-optics interferometers}

In a $n$-photon Boson Sampling experiment, the collection rate of significant output events scales unfavourably with the $n$-th power of the transmission of the optical setup. Hence, optical losses have to be reduced as much as possible, to enable experiments with larger and larger number of identical photons. In bulk optics, it may be natural to attribute a given amount of loss to the individual components, while propagation in free space is reasonably considered as lossless. On the contrary, in integrated optics some photon loss is intrinsic in waveguide propagation, especially in the curved parts, while components such as directional couplers may not introduce, because of their operation, specific additional loss. Therefore, in the design process of an integrated optical circuit, minimizing the propagation length is of paramount importance.

We discuss in the following different layouts for an integrated optics interferometer. We will evaluate in particular the footprint required  to achieve unitary transformations useful for a Boson Sampling experiment, and we will estimate how the physical size of the circuit scales with the number of modes. We will thus bring arguments in favour of our choice for a three-dimensional array of continuously-coupled waveguides, arranged on a triangular lattice.

\subsection*{Planar Clements interferometer}

Let us first determine the footprint of a planar circuit that performs an arbitrary unitary operation $U$ on $m$ optical modes, by means of a network of discrete phase shifters and beam splitters. We adopt the scheme proposed by Clements et al.~\cite{Clements:16}, which is indeed more compact than the earlier scheme by Reck et al.~\cite{Reck_1994}, and which has an optical depth of $m$ beam splitters. The optical depth is defined as the maximum number of such elements that a photon needs to pass through, being injected into an arbitrary input and exiting an arbitrary output port. 

\begin{figure*}
\centering
\includegraphics[scale=1]{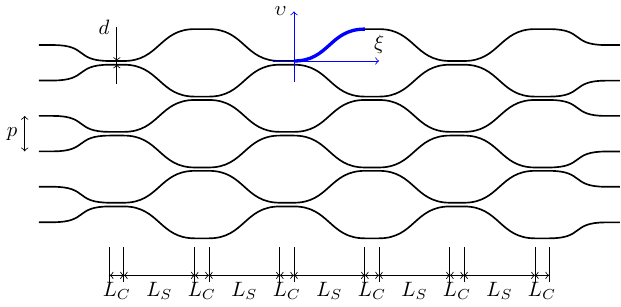}
\caption{\label{fig:clements} \textbf{Layout for universal interferometer.}. Schematic of an interferometer with $m=6$ modes, realized in integrated optics according to the Clements scheme. Black thick lines represent waveguide paths. Geometrical dimensions mentioned in the text are indicated. An S-bend curve, with its local coordinate system $(\xi,\upsilon)$ is also highlighted in blue color.}
\end{figure*}

In particular, we consider here an integrated-optics implementation of the Clements interferometer where beam-splitters are, in practice, waveguide directional couplers (see for instance the layout of a 6-modes interferometer in Supplementary Figure \ref{fig:clements}). It is in fact possible, with the femtosecond laser writing technology, to implement static networks of directional couplers with controlled phases and splitting ratios \cite{Crespi13boson}. We note that reconfigurable versions of the same scheme, equipped with thermo-optic phase-shifters \cite{dyakonov18}, require beam-splitters to be replaced by Mach-Zehnder interferometers. However, substitution of directional couplers with Mach-Zehnder devices further increases the length of the circuit and does not change how the length scales with the number of modes.

From Supplementary Figure \ref{fig:clements} we observe that, excluding the terminal waveguide segments, the overall length of an integrated Clements interferometer equals the sum of $m-1$ times the length $L_S$ of a waveguide S-bend, and $m$ times the length $L_C$ of the interaction region of the directional coupler (we may consider all couplers to have the same interaction length, as in Ref.~\cite{Crespi13boson}). 

The  technological platform chosen to implement the waveguide circuit determines the minimum curvature radius $R_{\mathrm{min}}$ that can be employed, above which additional losses due to curvature are negligible or at least tolerable. Once fixed such minimum radius, the length of the S-bend depends on the transverse elongation $h$ that it has to cover (the wider the elongation, the longer the S-bend). In fact, choosing a sinusoidal shape for the S-bend, the curve may be described by the equation:
\begin{equation}
\upsilon = h \sin^2 \left(\frac{\uppi}{2} \frac{\xi}{L_S}\right)
\end{equation}
in a suitable $(\xi,\upsilon)$ local coordinate system.
The minimum radius of curvature occurs at $\xi=0$ and $\xi=L_S$. By imposing that such minimum radius of curvature is precisely $R_{\mathrm{min}}$, the length of one S-bend in the $\xi$ direction is then given by the formula:
\begin{equation}
L_S = \frac{\uppi}{2} \sqrt{2 R_{\mathrm{min}} \, h}
\label{eq:LS}
\end{equation}

Directional couplers are formed by waveguides brought close one to the other at a small distance $d$ for a length $L_C$, and brought sufficiently far apart elsewhere, in order to quench the evanescent-field interaction, which typically decays exponentially with the distance.   In particular, interaction must have substantially decayed at the distance $p$ which is the pitch of the input and output waveguide segments. 
If $d \ll p$, the elongation of the S-bends inside the interferometric network can be approximated just as $h=p$. Therefore, we may consider:
\begin{equation}
L_S = \frac{\uppi}{2} \sqrt{2 R_{\mathrm{min}} \, p}
\end{equation}
In a practical case, taking the example of femtosecond-laser written circuits, $d$ may be in the order of a few microns and $p$ at least in the order of a few tens of microns.

The length of the interaction region $L_C$ of an individual directional coupler determines its reflectivity according to:
\begin{equation}
\mathcal R = \sin^2 (c \, L_C + \phi_0)
\end{equation}
where $c$ is the coupling coefficient between the waveguides which depends on the distance $d$, and $\phi_0$ takes into account coupling that occurs in the terminal parts of the S-bent waveguides. As a first approximation, we can consider $\phi_0=0$, and we can assume that the interferometer is designed as in Ref.~\cite{Crespi13boson}, hence:
\begin{equation}
L_C =  \frac{\uppi}{2c}
\end{equation}

Overall, the estimated length of the $m$-mode Clements interferometer will thus be given by:
\begin{align}
L_{\mathrm{Clem}} &= (m-1) \frac{\uppi}{2} \sqrt{2 R_{\mathrm{min}} p} + \frac{m \uppi}{2 c} = \notag\\
&=m \frac{\uppi}{2} \left(\sqrt{2 R_{\mathrm{min}} p} + \frac{1}{c}\right) - \frac{\uppi}{2}\sqrt{2 R_{\mathrm{min}} p}
\label{eq:Lclements}
\end{align}
It is worth noting that this length depends on the waveguide platform employed, which determines the order of magnitude for $R_{\mathrm{min}}$, $p$ and $c$. On the other hand, different platforms may be characterized by different amount of propagation losses for a given circuit length. 
In any case, for any waveguide platform, the circuit length given by Eq. \eqref{eq:Lclements} scales linearly with the number of modes.

For instance, in the case of femtosecond-laser written circuits it may be reasonable to take $R_{\mathrm{min}}$~=~30~mm, $p$~=~0.06~mm and a maximum exploitable $c$~=~1~mm$^{-1}$. This makes the two terms of the sum in Eq. \eqref{eq:Lclements} comparable, and results in a circuit footprint that increases of about 4.5~mm per each added mode to the interferometer. 

We could further note that, as mentioned, if we considered a fully reconfigurable circuit designed with the same architecture, each directional coupler should be replaced by a Mach-Zehnder interferometer, thus doubling the optical depth with respect to the case studied above. The precise dimensioning of this other circuit would deserve a separate and detailed discussion. As a rule of thumb, since the optical depth doubles, we could consider an increase of the circuit length of about 9~mm per each added mode, using analogous fabrication parameters.

\subsection*{Planar continuously-coupled interferometer}

Integrated-optics platforms enable the realization of interferometers that have no analogous with bulk components, where several waveguides are placed parallel one to the other, and are continuously coupled  by evanescent-field interaction. A general algorithm has still to be devised that teaches how to implement an arbitrary unitary transformation of the optical modes with this kind of devices. We will thus proceed with discussing the size constraints of these interferometers on the basis of heuristic arguments.  

\begin{figure}
\centering
\includegraphics[scale=1]{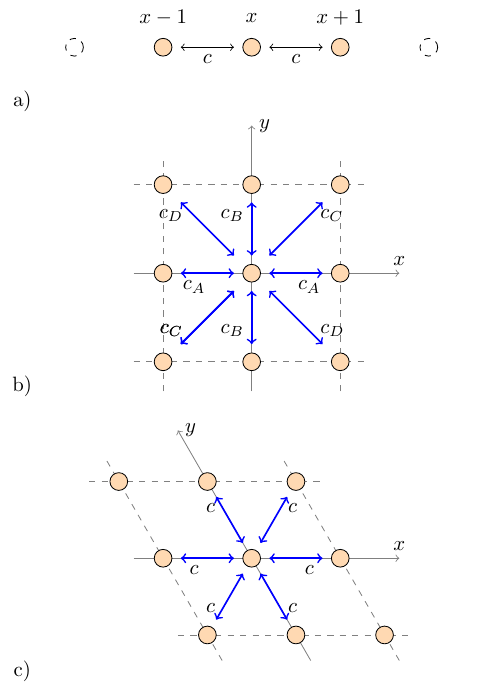}
\caption{\label{fig:couplings} \textbf{Geometries for continuously-coupled waveguide arrays}. (a) Scheme of the cross section of a planar array of identical waveguides. Optical modes (indexed with $x$) are coupled with a coupling coefficient $c$. (b) Concept scheme of the couplings considered in Eq. \eqref{eq:coupledModes2D} in the case of a Bravais lattice with primitive vectors parallel to the axis $x$ and $y$. In the general case, these axis may not be orthogonal. This scheme can be applied e.g. to a triangular lattice (c): in this case the directions $x$ and $y$ are separated by an angle of 120$^\circ$ and one can consider $c_A=c_B=c_C=c$, while $c_D=0$.}
\end{figure}

We can first consider an infinite array of identical optical waveguides placed parallel, on a plane, at the same distance $d$. Propagation of photons in the array can be described by coupled differential equations involving the destruction operators $\hat{a}_x$ of the optical modes. In the approximation of nearest-neighbour coupling, these equations read:
\begin{equation}
- \text{i} \frac{d \hat{a}_x}{dz} = c \hat{a}_{x-1} + c \hat{a}_{x+1}
\label{eq:coupledModes}
\end{equation}
where $z$ is waveguide direction, $c$ is the coupling coefficient which depends on the relative distance between two neighbouring waveguides, and $x$ is the mode index (see also Supplementary Figure \ref{fig:couplings}a).

We can look for a plane-wave solution of Eq. \eqref{eq:coupledModes} of the kind:
\begin{equation}
\hat{a}_x = \hat{\mathcal A} \, \text{e}^{\text{i} (\beta_z z + \beta_x x)}
\label{eq:planeWave}
\end{equation}
where $\beta_z$ and $\beta_x$ are components of the wavevector respectively parallel and transverse to the waveguide direction $z$. Note that in the parallel direction we measure the spatial propagation with $z$, that has the dimensions of a length, while for the transverse direction we use $x$, which is the waveguide index and is adimensional.
By substituting Eq. \eqref{eq:planeWave} into Eq. \eqref{eq:coupledModes} we derive a sort of dispersion relation \cite{eisenberg2000},
\begin{equation}
\beta_z = 2 c \cos (\beta_x)
\end{equation}
that governs light diffraction in such a discretized setting. 
In particular, an excitation characterized by a set of transverse components centered around $\beta_x$, will propagate  across the waveguides of the array with a group velocity: 
\begin{equation}
v_x = \frac{\partial \beta_z}{\partial \beta_x} = -2 c \sin \beta_x
\label{eq:vGroupLinear}
\end{equation}
Namely, after a distance $z$, the center of the wavepacket will have travelled across $\Delta x = v_x \,z$ waveguides. Note that there is a maximum achievable group velocity $|v_x|_{\mathrm{max}} = 2 c$. 

Let us consider now a photon that enters the array localized on a single waveguide mode. The state of this photon, being spatially pointlike (though in a discretized setting) contains all possible transverse components, including the ones travelling at $|v_x|_{\mathrm{max}}$.
The wavepacket will spread in the array along propagation in $z$; however, it will not be able to reach a waveguide that is placed $m$ positions apart in a propagation length smaller than:
\begin{equation}
L_{m} = \frac{m}{|v_x|_{\mathrm{max}}}= \frac{m}{2 c}
\label{eq:LmPlanar}
\end{equation} 
If we neglect boundary effects and we apply these considerations to a finite-size $m$-waveguide array, we can take this $L_{m}$ as an estimate for the minimum length that such array must have, so that photons entering in any given input port can interfere together in any output port. 

We note that this minimum length $L_m$ is shorter than the length of the Clements interferometer as expressed by Eq. \eqref{eq:Lclements}, if the same coupling coefficient $c$ is used in both circuits. However, a continuously-coupled waveguide array of length $L_m$ cannot reproduce the full set of unitaries allowed by the Clements architecture; a longer array with random modulations of the optical parameters may be needed to achieve similar purposes \cite{mural19,banchi17}. In addition, the length $L_m$ in Eq. \eqref{eq:LmPlanar} scales linearly with $m$ exactly as $L_{\mathrm{Clem}}$ in Eq. \eqref{eq:Lclements}.

\subsection*{3D continuously-coupled interferometer}

The femtosecond-laser-writing technology has the unique capability to inscribe waveguides at different depths below the glass surface, and thus to realize three-dimensional waveguide arrays with arbitrary cross-section \cite{pertsch04,szameit06,caruso2016maze}. Light propagation in these arrays is studied analytically by extending the coupled-equations formalism discussed above for the planar devices. In particular, Ref.~\cite{szameit07} reports general solutions for the coupling configuration schematized in Supplementary Figure \ref{fig:couplings}b, which is suitable to describe waveguides arranged according to a two-dimensional Bravais lattice with (possibly non-orthogonal) axes $x$ and $y$. Thus, Eq. \eqref{eq:coupledModes} is generalized as follows:
\begin{multline}
- \text{i} \frac{d \hat{a}_{x,y}}{dz} = c_A \left( \hat{a}_{x-1,y} + \hat{a}_{x+1,y}\right) + c_B \left( \hat{a}_{x,y-1} + \hat{a}_{x,y+1}\right) \\ + c_C \left( \hat{a}_{x-1,y-1} + \hat{a}_{x+1,y+1}\right) + c_D \left( \hat{a}_{x-1,y+1} + \hat{a}_{x+1,y-1}\right)
\label{eq:coupledModes2D}
\end{multline}
where $x$ and $y$ are waveguide indices on the two directions of the Bravais lattice, $c_A$ and $c_B$ are coupling coefficients between neighbouring sites on these two separate directions, $c_C$ and $c_D$ take into account also diagonal coupling.
Plane-wave solutions of Eq. \eqref{eq:coupledModes2D} take the form:
\begin{equation}
\hat{a}_{x,y} = \mathcal A \, \text{e}^{\text{i} (\beta_z z + \beta_x x + \beta_y y)}
\label{eq:planeWave2D}
\end{equation}
which accounts for distinct $\beta_x$ and $\beta_y$ on the two Bravais directions.

The case of an infinite square lattice, with negligible diagonal coupling, corresponds to $c_A = c_B = c$ and $c_C=c_D=0$. In this case, the following dispersion relation is retrieved:
\begin{equation}
\beta_z = 2c \left(\cos \beta_x + \cos \beta_y \right)
\end{equation}
The group velocity of a wavepacket propagating only along the $x$ or $y$ direction takes an expression analogous to Eq. \eqref{eq:vGroupLinear}, with the same maximum value $|v_x|_{\mathrm{max}} = |v_y|_{\mathrm{max}} = 2 c$.

Therefore, if we compare a planar array and a three-dimensional array with a square-lattice waveguide arrangement, having the same coupling strength, we observe that light can propagate transversally across the waveguides of the former with the same maximum velocity experienced in the latter, along the $x$ or $y$ direction taken separately. However, in the case of point-like excitation of the square lattice, light travels simultaneously in both directions thus spreading to a quadratically larger number of waveguides.   

We can transfer these considerations to a finite-size array, again neglecting boundary effects for simplicity. In a finite-size square lattice with a total of $m$ waveguides, the maximum waveguide indices along the two directions $x$ and $y$ are proportional to $\sqrt{m}$. Therefore, in order to spread on the full area of the array, a wavepacket will have to propagate along a transversal direction across a number of waveguides proportional to $\sqrt{m}$. It is thus reasonable to assume that the minimum array length, required for allowing a photon injected into an arbitrary input waveguide to reach an arbitrary output waveguide, is:  
\begin{equation}
L_{m} = B \frac{\sqrt{m}}{|v_{x,y}|_{\mathrm{max}}}= \frac{B}{2 c} \sqrt{m}
\label{eq:LmSquare}
\end{equation} 
where $B$ is some constant. We note that here $L_{m}$ scales more favourably with respect to the case of the planar array, being proportional to $\sqrt{m}$ and not to $m$.

An even faster light spreading can be reached in triangular arrays (see Supplementary Figure \ref{fig:couplings}c), as the ones we adopted in the experiments described in the Main Text. In this case $c_A = c_B = c_C = c$ while $c_D=0$, and the dispersion relation reads \cite{szameit06, szameit07}:
\begin{equation}
\beta_z = 2c \left(\cos \beta_x + \cos \beta_y + \cos (\beta_x + \beta_y)\right)
\end{equation}
Note that, due to the definition of the Bravais vectors, here $x$ a $y$ are non-orthogonal directions, oriented with a relative angle of 120$^\circ$.  The group velocity along the $x$ or $y$ direction is given by:
\begin{equation}
v_{x,y} = \frac{\partial \beta_z}{\partial \beta_{x,y}} = -4 c \sin \beta_{x,y}
\label{eq:vGroupTria}
\end{equation}
with a maximum value $|v_x|_{\mathrm{max}} = |v_y|_{\mathrm{max}} = 4 c$, which is twice that retrieved for linear or square lattices.

Scaling properties of this minimum length $L_m$, namely the dependence on $\sqrt{m}$, are analogous to the ones discussed for the square-lattice case and resulting in Eq. \eqref{eq:LmSquare}. However, in the triangular-lattice case $L_{m}$ should roughly halve its value with respect to the former case. 

All considerations made up to now regard homogeneous arrays, in which coupling coefficients are uniform across the array and along the waveguide direction. In addition, the evaluated lengths $L_m$ are only related to the spreading of the light to all output waveguides, with no constraints on the unitary transformation provided by the circuit. As a matter of fact, homogeneous arrays produce highly symmetric transformations, which may not preserve the complexity of the Boson Sampling problem \cite{mural19}. 

In our work, we are indeed introducing static randomness in the continuously-coupled interferometer by taking slightly different coupling coefficients $c$ between different waveguides, namely by varying the relative interwaveguide distances, and by further modulating them along $z$. In addition, we use an array length that is more than twice the estimated $L_{m}$ with our experimental parameters, in order to promote further mixing of the light across the waveguide array. In fact, in the case of our device, with $m=32$ and an average coupling coefficient $c \sim 0.2\;\mathrm{mm}^{-1}$, $L_{m}$ is in the order of 15~mm while the waveguide array reaches the length of 36~mm. The considerations about fast light spreading in the arrays, presented in this Section, have guided us in the choice of the triangular array geometry with respect to other possible interferometer layouts.

We may note that previous theoretical works have shown that a random time-modulation of the site energies, in a quantum walk on a one-dimensional chain of sites, is able to provide Haar-random transformations of the input states in the long-time scale \cite{banchi17,mural19}. In a photonic setting, such a quantum walk can be implemented by a photon propagating in planar waveguide arrays, where the propagation constant of the waveguides is modulated randomly along the $z$ direction, and in a different way in each waveguide. A random transformation would then be achieved with a sufficiently long device. In our experiments, fluctuations of the waveguide positions in the cross-section of the array provide a static random modulation of the coupling coefficients, while non-uniform modulations in the propagation constants are provided in a reconfigurable way through the thermo-optic effect, by employing the resistive micro-heaters patterned on the chip surface. The achieved randomness has been investigated experimentally, as described in the Main Text.

\subsection*{Fan-in and fan-out sections}

A waveguide interferometer reasonably needs to be coupled, at the input and output ports, with optical fibers. To connect several fibers in parallel to the same facet of the optical chip, commercial fiber blocks or fiber arrays are a convenient choice. These components typically contain the desired number of fibers, arranged on a line at a fixed pitch $p_F$, whose precise alignment is guaranteed by means of V-grooves fabricated lithographically.

In the case of the Clements interferometer, one may choose to design the pitch $p \equiv p_F$, so that fibers can be coupled in a direct fashion. However, in the general case, and especially in continuously-coupled waveguide arrays, the pitch of the waveguides is different, and likely quite smaller, than the pitch of the fiber array. In addition, in case of three-dimensional arrays, waveguides are not even arranged along a single line. In such circumstances, fan-in and fan-out sections need to be added to the circuit, which are composed of S-bends that bring the waveguides at the correct spacing and with the correct arrangement to be interfaced with the fibers.

It is important to study how the length scales also for these parts of the circuit. In fact, we should at least check that advantages in compactness gained in the interferometer section are not vanished by long fan-in and fan-out sections.

As previously discussed, the length of sinusoidal S-bends is governed by Eq. \eqref{eq:LS}. If the fan-in or fan-out is judiciously designed, it is reasonable to assume that the maximum lateral elongation will be smaller than half the width of the fiber array, namely  $h_{\mathrm{max}} < p_F \cdot (m-1)/2$.
This constrains the length of the fan-in or fan-out sections to:
\begin{equation}
L_{F} = \frac{\uppi}{2} \sqrt{2 R_{\mathrm{min}} h_{\mathrm{max}}} < \frac{\uppi}{2} \sqrt{(m-1) R_{\mathrm{min}} p_F }
\label{eq:LFlinear}
\end{equation}
We note that this bound for $L_F$ scales with $\sqrt{m}$ as $m$ grows larger. Therefore, as the number of modes increases, the length of these sections tends to be less relevant than the length of a Clements interferometer, which scales as $m$.
On the other hand, it may share a similar scaling law as a three-dimensional continuously-coupled interferometer.

An improved compactness may be gained if the $m$ fibers to be coupled are arranged on a two-dimensional grid, as in the experiment described in the Main Text. In this case, the maximum size of the fiber array cross-section, in either dimension, scales as $\sqrt{m}$. This maximum size determines also the maximum elongation of the S-bend. Applying Eq. \eqref{eq:LS}, it follows that the length of the fan-in or fan-out section here scales as $\sqrt[4]{m}$. This further contributes to limiting the device footprint and, hence, its optical insertion losses.


\section{Reconstruction algorithm}
\begin{figure}[t]
    \centering
    \includegraphics[width=0.9\columnwidth]{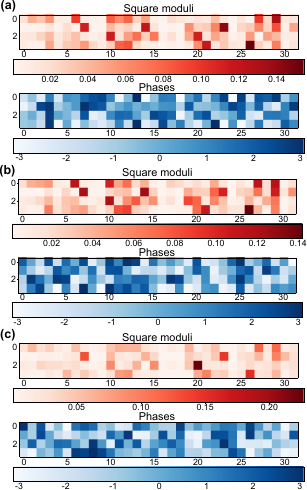}
    \caption{\textbf{Reconstruction of the unitary matrix.} Examples of the squared moduli and phases retrieved through the algorithm. Each case corresponds to different configuration of the circuit. The rows represent the four input of the photons and the 32 columns are the output modes of the device.}
    \label{fig:unitary_rec}
\end{figure}
A fundamental point for the characterization of reconfigurable photonic chip and for Boson Sampling itself is the reconstruction of the associated unitary matrix. In the case of continuous-coupling devices it is not trivial to find an analytical model linking the elements of the unitary matrix to the internal architecture of the device. For this reason, we model the sub-matrix elements in the most general way, i.e as complex numbers $U_{lm}=\rho_{lm} \text{e}^{i \phi_{lm}}$ identified by the modulus $\rho_{lm}$ and the phase $\phi_{lm}$. We start by considering two quantities related to twofold experiments. The first is the probability to detect two distinguishable particles in the output $i,j$, when the particles are injected from the input $h,k$
\begin{equation}
    a_{ij}^{hk}=\rho_{ih}^2 \rho_{jk}^2+\rho_{jh}^2 \rho_{ik}^2
    \label{eq: plateau}
\end{equation}
We observe that the values of $a_{ij}^{hk}$ correspond to a direct measurement of the unitary matrix moduli. The second quantity is the visibility of the Hong-Ou-Mandel (HOM) dip, defined as the difference between $a_{ij}^{hk}$ and the probability of detecting two indistinguishable particles divided by $a_{ij}^{hk}$:
\begin{align}
\label{eq:visibility}
    V_{ij}^{hk} &= \frac{a_{ij}^{hk} -|U_{ih}U_{jk}+U_{jh}U_{ik}|^2}{a_{i,j}^{h,k}} = \\ 
    & = -\frac{2 \rho_{ih}\rho_{jk}\rho_{jh}\rho_{ik} }{a_{ij}^{hk}}\cdot \cos ( \phi_{ih} + \phi_{jk}-\phi_{jh}-\phi_{ik} ) \notag
    \end{align}
The visibility $V_{ij}^{hk}$ is sensitive to the difference among the phases of matrix elements. The $496$ values of the $a_{ij}^{hk}$ and $V_{ij}^{hk}$ for a fixed input pair $h,k$ can be retrieved from the experimental data via different procedures. The $a_{ij}^{hk}$ can be calculated from the intensity distribution of classical light or from the distribution of heralded single-photon injected in $h$ and $k$. The visibilities can be derived from direct measurements of the two photons probabilities in two different positions of the delay lines. Alternatively, the interpolation of the data resulting from a complete scan of the position $x$ of the delay lines in the HOM dips with a Gaussian function
\begin{equation}
f(x)=a_{ij}^{hk}\biggl(1+V_{ij}^{hk}\text{e}^{-\frac{(x-x_0)^2}{2\sigma^2}}\biggr)
\end{equation}
provides an estimation of the quantities under investigation.

We find the moduli of the matrix elements by minimizing the $\chi^{(2)}$ quantity between the measured $a_{ij}^{hk}$ and the right side of Eq. \eqref{eq: plateau}. For the phases we exploited an analytical algorithm proposed in Ref. \cite{laing2012superstable} to solve the equation \eqref{eq:visibility}. One limitation of this algorithm is that it is not robust with respect to experimental imperfections. For this reason, we use such solution only as a starting point for a further minimization to find the phases $\phi_{lm}$ by fixing the moduli to the values retrieved from the first minimization. The quantity to optimize is 
\begin{multline}
  \chi^{(2)} = \sum_{h,k} \sum_{i,j} \bigg [a_{ij}^{hk}\big(1+V_{i,j}^{hk}\big)+\\
  -|U_{ih}U_{jk}+U_{jh}U_{ik}|^2\bigg]^2/ (\epsilon_{ij}^{hk})^2 
   \label{eq:visibility2}
\end{multline}
where $\epsilon_{ij}^{hk}$ are the experimental error associated to the quantity $a_{ij}^{hk}\big(1+V_{i,j}^{hk}\big)$. Furthermore, this last minimization enables us to reconstruct the $3\times 32$ and $4\times 32$ sub-matrices from a subset of the possible input pairs necessary to evaluate the phases values. In fact, using such pool of data, the analytical algorithm provides two or more solutions that differ only in the sign of the phases. The minimization discriminates which solution provides a better agreement with the experimental data.

In Supplementary Figure \ref{fig:unitary_rec} we report some examples of $4\times32$ sub-matrix reconstructed using the method described in this section.


\section{Unitary transformations}
\begin{figure}[t]
    \centering
    \includegraphics[width=\columnwidth]{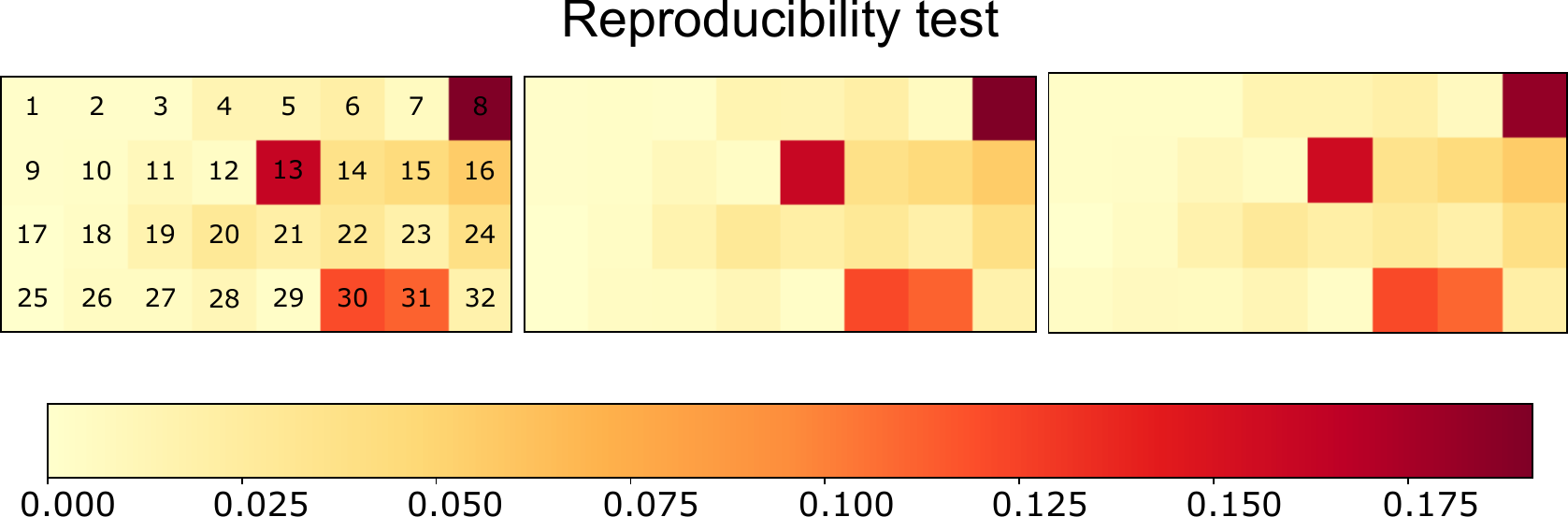}
    \caption{\textbf{Reproducibility of the implemented transformations.} We report the intensity patterns of the output modes for a fixed circuit configuration when a single heralded photon is injected in one input port. The average similarity between two of the three maps is 99.8$\%$. The enumeration on the left map denotes the respective output mode's intensity. The other two maps follow the same enumeration.}
    \label{fig:reprod}
\end{figure}
\begin{figure}[t]
    \centering
    \includegraphics[width=\columnwidth]{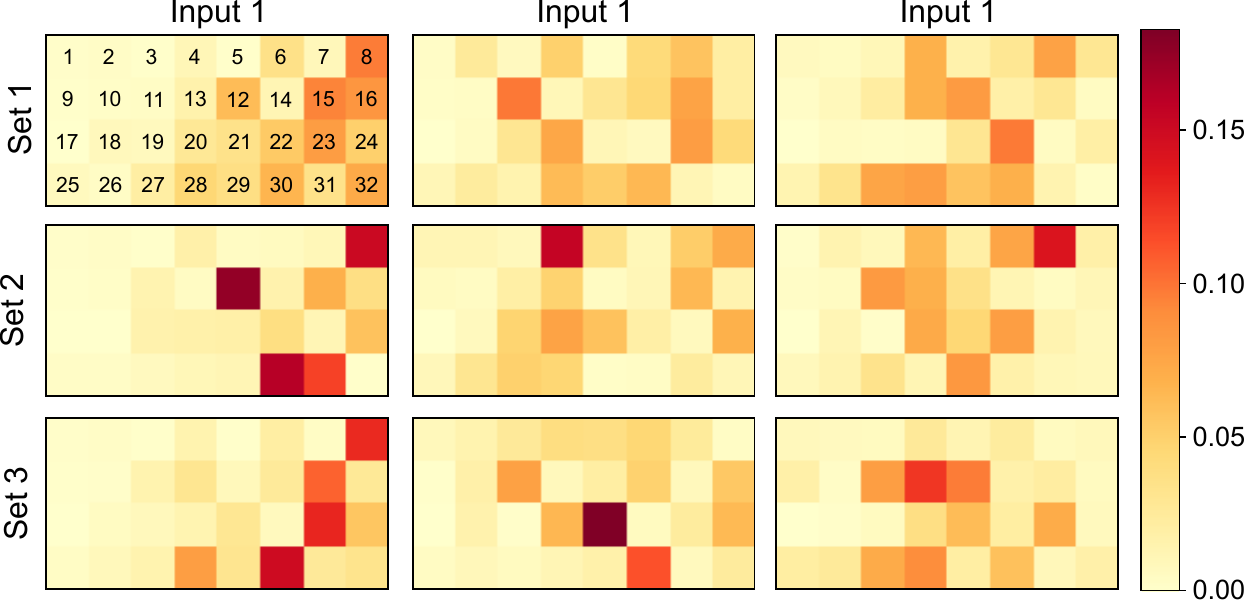}
    \caption{\textbf{Reconfigurability of the photonic chip.} Here we report the intensity patterns of the 32 output modes (which are illustrated accordingly to the labelling in the top left pattern) for {3 sets of circuit configurations and for 3 different input ports}. Thus, we can observe the capability of the chip to implement different unitary transformations while varying the applied currents.}
    \label{fig:recon}
\end{figure}
In this section we describe the preliminary tests performed with the photonic chip regarding the ability to cover a large amount of unitary transformations and to reproduce a given distribution in the outputs over time.  
In order to check the last feature of the device, 
we injected a single heralded photon in a fixed input port of the chip, and we collected the output intensities of the 32 modes. Thus, we obtained the squared moduli of a single column of the implemented unitary. We repeated the measurement three times in different days, preserving the same setting for the applied currents. In Supplementary Figure \ref{fig:reprod} we report the measured output distributions. We quantified the reproducibility through the similarity between the three patterns, whose average is $99.8\%$. 

For what concerns the reconfigurability of the chip, namely the capability to implement different transformations on the input state, we performed the same measurement of the stability test. In this case we varied the applied currents in the circuit through a uniform sampling of the dissipated electrical power. The collected output distributions are included in Supplementary Figure \ref{fig:recon}.

{
In {} Fig. 3b-c {} we have {reported} the distributions of the squared moduli and the phases of the various experimental matrices and compared them with likewise matrices extracted according to the Haar distribution. There are slight discrepancies between the experimental and theoretical distributions. In this section we investigate a simplified model to identify the experimental imperfections that generate such deviations from the expected distributions.  
The histogram of the squared moduli in Fig. 3b {} is more peaked towards zero than the theoretical one. This is {likely} due to errors in the estimation of input and output losses. To {support this hypothesis}, we performed a numerical simulation in which an error of at most 10\% on the estimated values of the losses was inserted in the Haar random matrices. The results are shown in Supplementary Figure \ref{fig:simulations}a. The green distribution that displays losses is in good agreement with the experimental one reported in the main text. For what concerns Fig. 3c {}, we observe {} a slight peak around zero in the experimental phase{} distribution. In this case, such discrepancy {can be attributed} to residual correlations between the phases of the neighboring waveguides and to the reconstruction method of unitary matrices. {To study this aspect we have performed another} numerical simulation that exploits a simplified model of our device. We consider the case in which the first neighbouring couplings are static along $z$ and {} the whole surface is uniformly heated. This produces a linear gradient of the temperature at different {depths in} the sample, which in turn generates correlated changes in the propagation constants and thus in the matrix phases. Note that this condition is quite far from the experiment in which each heater has been controlled independently. The red curve in Supplementary Figure \ref{fig:simulations}b corresponds to {the resulting phase} distribution. {Interestingly, this distribution} is completely flat as for the case of Haar random  matrices. This {highlights the fact} that a flat distribution of the unitary {matrix} phases alone is not a sufficient proof for the sampling from the Haar distribution. The green histogram{, on the other hand,} corresponds to the same {ensemble of matrices, multiplied both at left and at right by diagonal matrices of unit-valued complex elements (equivalent to phase shifters placed at the inputs and outputs of the device), in such a way that} the phases of the first columns and rows {are} set to zero. The latter {zero-valued phases} were not included in the histogram distribution. Such scenario reproduces what we did in the reconstruction algorithm of the experimental matrix; in fact, the measured HOM visibilities are not sensitive to the input and output individual phases and this allows us to set them to 0 in the corresponding first row and column of the matrix. This procedure generates the slight concentration of the phases in zero, as shown in Supplementary Figure \ref{fig:simulations}b.}

\begin{figure}[t]
    \centering
    \includegraphics[width=0.8\columnwidth]{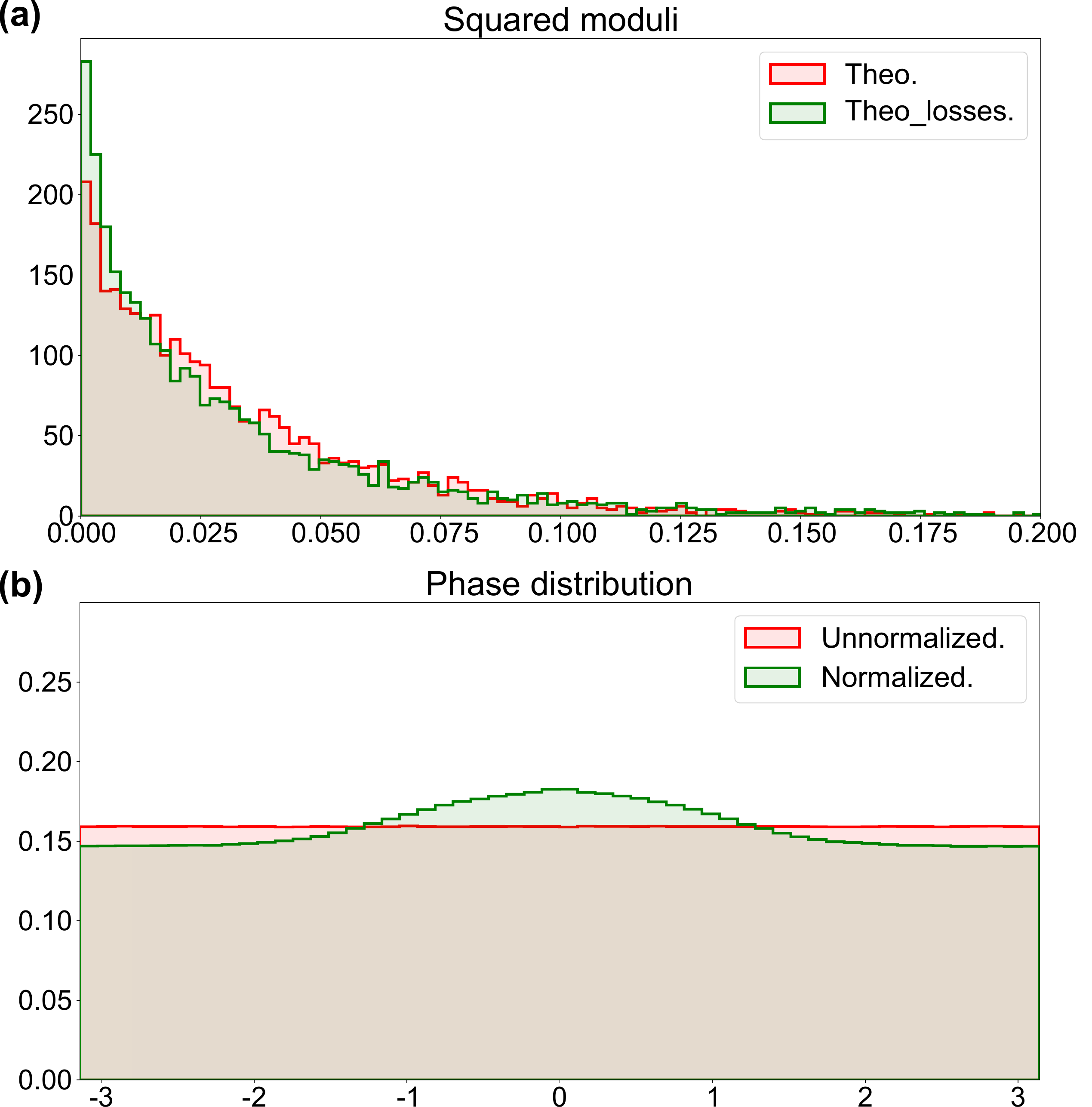}
    \caption{{\textbf{Discrepancy from the Haar random matrices distributions.} (a) Comparison between the squared-moduli distributions of 15 $[3\times32]$ ideal sub-matri{ces} from Haar-random-extracted unitary transformations (red) and the same sub-matri{ces} with {randomly modulated} insertion losses in the input and output stages (green). The lossy case reproduces the experimental distribution reported in Fig. 3b in the main text. (b) Numerical simulation for the sub-matrix phases distributions in the presence of correlations due to uniform heating of device{} surface. In red the resulting phases distribution and in green the same phases expressed by using the phases of the first column and row as reference. In this second case the distribution is not flat.}}
    \label{fig:simulations}
\end{figure}


\section{Single-photon source}
In this section we illustrate the adopted model used for the four-photon states generated by the spontaneous parametric down-conversion in a BBO source. The double-pair emission by a nonlinear crystal can be considered as two independent emission processes by likewise sources. Then, the state produced by each source is $\ket{\psi_{ij}}\sim \ket{00}\bra{00}+g_{ij}\ket{11}\bra{11} + {g_{ij}}^2\ket{22}\bra{22}+\dots$, where $g_{ij}$ is the nonlinear gain for the source emitting in the modes $(ij)=\{(1,2), (3,4)\}$ (see Supplementary Figure \ref{source}). For a complete description of the state, the losses, labelled by $\eta_i$ in Supplementary Figure \ref{source}, were taken into account. In our experimental setup the corresponding losses are associated to the coupling in single-mode fibers (particularly in delay-lines). Then, the input state resulting from the product $\rho^{\mathrm{in}} \sim \ket{\psi_{1,2}} \cdot \ket{\psi_{3,4}}$, includes the contributions with different number of photons, weighted by coefficients that depend from $g_{(ij)}$ and $\eta_i$. The state, post-selected by measuring four-fold coincidence, has the following form when expressed in the occupation numbers $\ket{n_4, n_1, n_2, n_3}$, i.e through the number of photons in the corresponding input mode
\begin{align}
   \rho^{4-\mathrm{photon}} & \sim \frac{1}{\alpha + \beta +\gamma} \big (\alpha \ket{1111}\bra{1111}+ \notag \\
    & +\beta\ket{2002}\bra{2002} + \gamma\ket{0220}\bra{0220} \big ),  
    \label{eq:input_state}
\end{align}
with
\begin{align}
    \alpha & = g_{12} g_{34} \eta_1 \eta_2 \eta_3 \eta_4 \\ \notag
    \beta & = g^2_{34} \eta_3^2 \eta_4^2 \\ \notag
    \gamma & = g^2_{12} \eta_1^2 \eta_2^2. 
\end{align}
The above coefficients can be retrieved from direct measurements of two-fold coincidences which are related to the contributions $\ket{1001}$ and $\ket{0110}$ in $\rho^{\mathrm{in}}$. In fact, the ratio between these counts is $R = \frac{g_{34} \eta_3 \eta_4}{g_{12} \eta_1 \eta_2}$. Then, the three coefficients have a straightforward expression in $R$, namely $\gamma=1$, $\alpha=R$ and $\beta = R^2$.

In the 3-photon Boson Sampling experiment we post-selected the contribution $\ket{1111}$ by detecting one photon in mode 4 and the other three photons in the chip's outputs. In this case, the input state does not depend from the relative weights among the various contributions in Eq. \eqref{eq:input_state}. The estimation of $\alpha$, $\beta$ and $\gamma$ is pivotal in the 4-photon experiments, where the whole state is injected in the optical circuit. We have characterized all coefficients by measuring the $R$ parameter from the two-fold coincidences before the chip. This preliminary characterization of the source was necessary for the validation of the 4-photon samples.

We conclude this section by providing some information about the generation rate of the source. The rate of twofold generation was $\sim 18$ kHz in the input modes $1,2$ and $\sim 16$ kHz in $3,4$. For what concerns the Boson Sampling experiment, we had an average rate of $\sim 300$ events per hour for the threefold case and $\sim 60$ for the fourfold events after the chip.


\begin{figure}
    \centering
    \includegraphics[width=0.5\columnwidth]{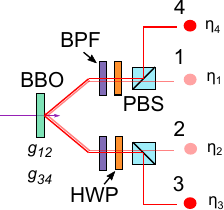}
    \caption{\textbf{Scheme for the double-pair emission process.} The source is split in two independent double pairs emissions characterized by two nonlinear gain $g_{ij}$. Each source generates photon in pair of optical modes. The latter are identified by propagation losses $\eta_i$. \textbf{Legend:} BB0 - Beta-Barium-Borate crystal; BPF - Band Pass Filter; HWP - Half Wave-plate; PBS - Polarizing Beam-splitter.}
    \label{source}
\end{figure}


\section{Boson Sampling validations}

In the experiment we have benchmarked the integrated device by performing several Boson Sampling experiments with 3- and 4-photon states. We repeated the measurements for 10 different configurations of the optical circuits for the 3-photon case, and further three configurations for the 4-photon state in Eq. \eqref{eq:input_state}. We adopted likelihood ratio tests to assign the data to a given hypothesis. First, the data were validated against the uniform sampler \cite{Aaronson14}. This algorithm requires the estimation of the quantifier $\mathcal{P} = \prod_i \sum_j |U_{ij}|^2$ where the index $i$ labels the modes in which photons are detected, the index $j$ the input modes and $U$ the unitary matrix representing the circuit. The counter $W$ initialized to zero is updated after the measurement $k$ according to the following rule
\begin{equation}
W_k =
    \begin{cases}
    W_{k-1}+1 & \text{if $\mathcal{P}\geq \left(\frac{n}{m}\right)^n$}\\
    W_{k-1}-1 & \text{if $\mathcal{P} < \left(\frac{n}{m}\right)^n$},\\
    \end{cases}
\end{equation}
where $n$ and $m$ are the number of photons and modes in the optical circuit.
The intuition behind this method is that the quantifier $\mathcal{P}$ reflects somehow the probability to observe the outcome $k$. If such quantity is greater than the uniform probability is plausible that the event was sampled from a nontrivial distribution. The second test applied to the Boson Sampling data regards the validation against the distinguishable particles \cite{Spagnolo14}. In this case the quantifier is the ratio between the probability $q = |\text{Per}\,U_{(ij)}|^2$ to detect indistinguishable particles, and the probability $d =\text{Per}\,|U_{(ij)}|^2$ to detect distinguishable particles in the set of output modes $j$ given the input modes labelled by $i$. $U_{(ij)}$ stands for the sub-matrix identified by the input labels $i$ and output labels $j$ and Per is the matrix permanent. By defining $\mathcal{L}=\frac{q}{d}$, the counter $C$ is updated after each outcome $k$ from the Boson Sampling as
\begin{equation}
C_k =
    \begin{cases}
    C_{k-1}+1 & \text{if $\mathcal{L}\geq 1 $}\\
    C_{k-1}-1 & \text{if $\mathcal{L} < 1$},\\
    \end{cases}
\end{equation}

\begin{figure}[t]
    \centering
    \includegraphics[width = 0.8\columnwidth]{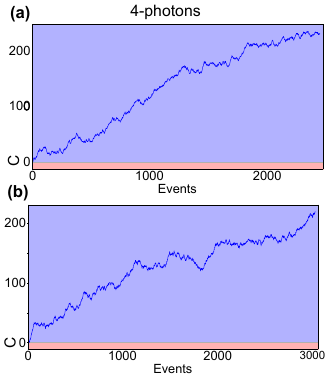}
    \caption{\textbf{4-photon experiments validation.} We report the validations against the distinguishable particle hypothesis for the state in Eq. \eqref{eq:input_state}. These further 4-photon experiments were not reported in the main text.
    }
    \label{fig:val_4ph}
\end{figure}

\begin{figure}[ht!]
    \centering
    \includegraphics[width = 0.8\columnwidth]{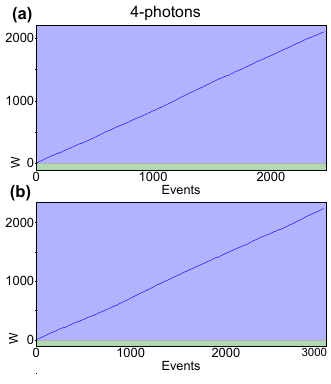}
    \caption{\textbf{4-photon experiments validation.} We report the validations against the uniform hypothesis for the state in Eq. \eqref{eq:input_state}. These further 4-photon experiments were not reported in the main text.
    }
    \label{fig:val_4ph_unif}
\end{figure}

\begin{figure*}[ht!]
    \centering
    \includegraphics[scale = 0.87]{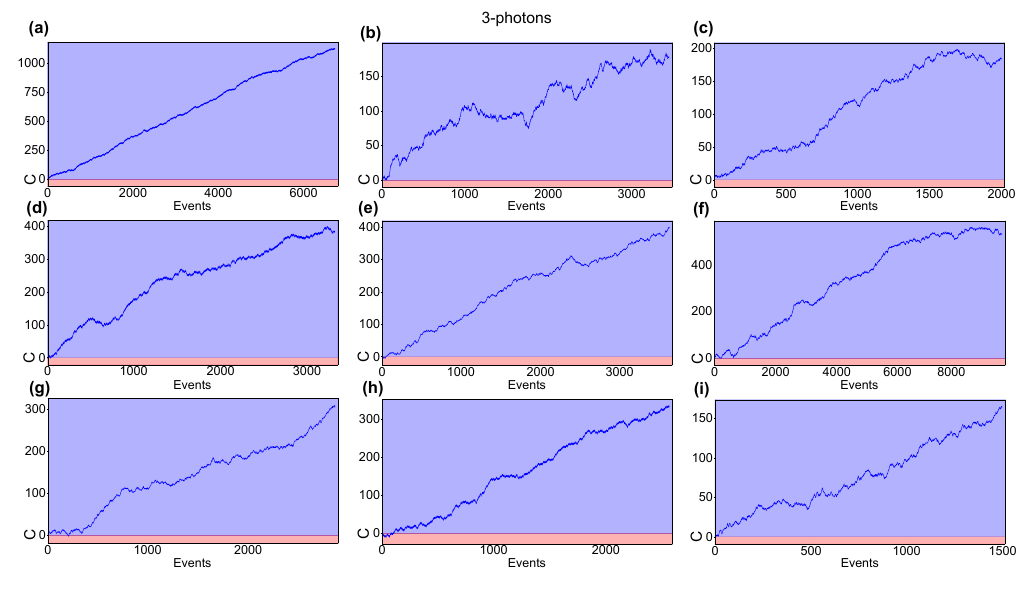}
    \caption{\textbf{3-photon Boson Sampling validations against the distinguishable particle sampler.} We report the complete set of Boson Sampling experiments for the 3-photon case states. The plots report the validation against the distinguishable samplers for different setting of the optical circuits. The validation is successful in all the nine cases.}
    \label{fig:val_3ph}
\end{figure*}

\begin{figure*}[ht!]
    \centering
    \includegraphics[scale = 0.87]{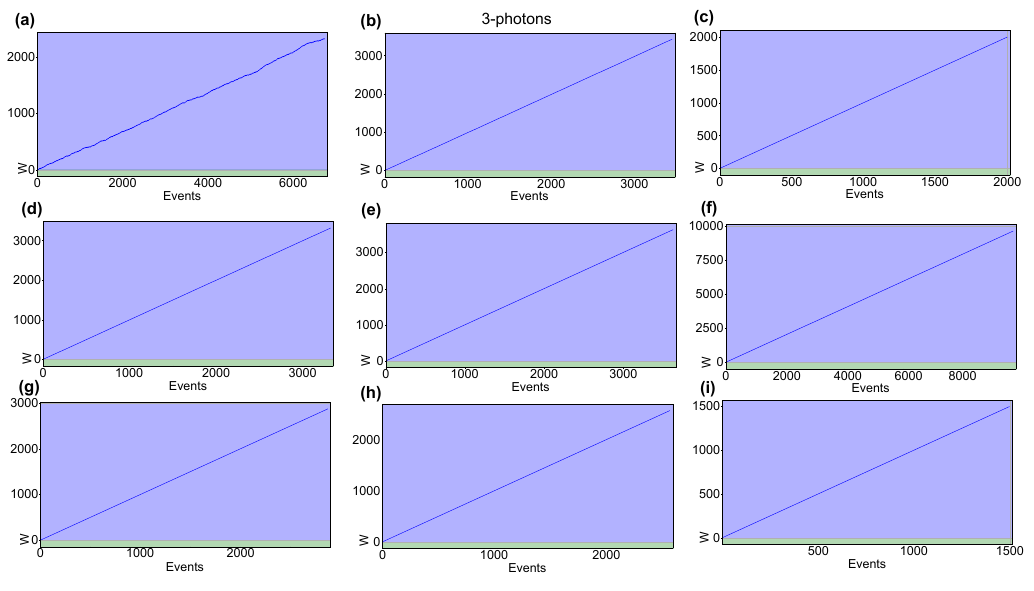}
    \caption{\textbf{3-photon Boson Sampling validations against the uniform distribution.} We report the complete set of Boson Sampling experiments for the 3-photon case states. The plots report the validation against the uniform sampler for different setting of the optical circuits. The validation is successful in all the nine cases.}
    \label{fig:val_3ph_unif}
\end{figure*}

Note that the expressions of $p$ and $q$ are related to the collision-free 
subspace accessible in the reported experiment. Furthermore, both quantifiers $\mathcal{P}$ and $\mathcal{L}$ depend from the the element of the matrix $U$ representing the interferometer. In the main text we have shown how a incorrect reconstruction of the unitary matrix affects the outcome of the test. 

In Supplementary Figures \ref{fig:val_4ph}-\ref{fig:val_4ph_unif} we report the validations tests performed for the two 4-photon experiments not included in the main text.
In Supplementary Figures \ref{fig:val_3ph}-\ref{fig:val_3ph_unif} we have reported the nine 3-photon Boson Sampling validations whose slope values are reported in Fig. 4f of the main text.


